\documentclass[reprint,superscriptaddress,amsmath,amssymb,aip]{revtex4-1}

\usepackage{graphicx}
\usepackage{dcolumn}
\usepackage{bm}
\usepackage{color, colortbl}
\usepackage{multirow}
\usepackage{amsmath}
\usepackage[version=4]{mhchem}
\usepackage{enumitem}
\usepackage{graphicx}
\usepackage{caption}
\usepackage{stfloats} 
\usepackage{lipsum}
\definecolor{dark-green}{rgb}{0.1,0.5,0.1}

\usepackage{caption}

\begin{document}

\title{Magnetoelastic coupling descriptor for high-throughput ab initio search of magnetocaloric materials}

\author{Debajit Chakraborty}
\affiliation{Department of Physics and Astronomy, University of Nebraska at Kearney, Kearney, Nebraska 68849, USA}

\author{Kirill D. Belashchenko}
\affiliation{Department of Physics and Astronomy and Nebraska Center for Materials and Nanoscience, University of Nebraska-Lincoln, Lincoln, Nebraska 68588, USA}

\author{Aleksander L. Wysocki}
\email{wysockia@unk.edu}
\affiliation{Department of Physics and Astronomy, University of Nebraska at Kearney, Kearney, Nebraska 68849, USA}

\begin{abstract}
We use Landau theory of phase transitions to design a magnetoelastic coupling descriptor that can identify magnetic materials on the brink of a first-order transition, where a strong magnetocaloric effect (MCE) can arise. The descriptor can be computed from electronic structure calculations and requires structural optimization in the paramagnetic state, which we model using special quasi-random collinear spin configurations. We first evaluate the descriptor for a set of known magnetocaloric materials and identify compounds in which the MCE is driven by magnetoelastic coupling. We then apply the descriptor in a high-throughput \emph{ab initio} screening of magnetic L1$_2$ compounds. This search identifies the L1$_2$ cubic phases of Mn$_3$Ge and Mn$_3$Sb as materials with strong magnetoelastic coupling and potential for a large magnetocaloric response.
\end{abstract}

\date{\today}

\maketitle

\section{Introduction}

The magnetocaloric effect (MCE)\cite{Pecharsky1999,Tishin2003} involves heat exchange between a ferromagnetic (FM) material and its environment in response to a change in the applied magnetic field. MCE has a broad range of applications, including magnetic heat pumping, refrigeration \cite{Franco2018}, energy conversion \cite{Kitanovski2020}, and magnetic hyperthermia \cite{Pimentel2018}. Among these, magnetic refrigeration plays a particularly important role in advancing efficient, solid-state cooling technologies, offering an environmentally friendly alternative to conventional vapor compression systems.

Although numerous materials exhibiting a MCE have been discovered \cite{Gottschall2019,Zhang2024}, the majority contain elements that are expensive (such as Rh, Ge, and In), toxic (including As and P), or critical in supply (like Gd and Y). Furthermore, integrating these materials into practical devices often demands complex, time-consuming synthesis and processing techniques \cite{Smith2012}. These challenges significantly hinder the large-scale industrial adoption of magnetic refrigeration. A successful transition from conventional vapor-compression-based cooling and heat pumping systems to the more energy-efficient and environmentally friendly magnetocaloric technology depends on expanding the availability of high-performance, cost-effective, and easily manufacturable magnetocaloric materials that operate near room temperature.

High-throughput (HTP) electronic structure calculations offer a promising approach for identifying new magnetocaloric compounds\cite{Zarkevich2018,Garcia2020,Shen2021,Fortunato2023,Romero2023,Fortunato2024}. However, a major challenge in such HTP searches is that the key magnetocaloric figure of merit, the isothermal entropy change in response to an applied magnetic field, is not directly accessible through standard electronic structure methods. To overcome this limitation, it is essential to develop an appropriate descriptor that can be used to screen potential magnetocaloric materials. This descriptor should capture the underlying physical mechanisms responsible for MCE, while also being computable from quantities readily obtainable via electronic structure calculations.

According to the Maxwell relation, the isothermal entropy change is related to the temperature derivative of the magnetization. As a result, the MCE is most pronounced near magnetic phase transitions, where the temperature dependence of the magnetization is steepest. In particular, the MCE can be exceptionally strong at first-order FM transitions due to the abrupt change in magnetization with temperature. A common mechanism for a first-order magnetic transition is a coupled magnetostructural instability, in which the magnetic and structural transitions occur simultaneously. MnAs is a canonical example, where the low-temperature FM hexagonal phase transforms first order into a paramagnetic (PM) orthorhombic phase \cite{MnAs}, generating a giant entropy change\cite{MnAs2}. However, the magnetic hysteresis and discontinuous volume changes often associated with such strongly first-order transitions pose significant challenges for practical magnetic refrigeration \cite{Scheibel2018}. Therefore, materials that exhibit weakly first-order transitions (or lie close to a tricritical point), not necessarily driven by a magnetostructural instability, are of particular interest for developing efficient and reliable magnetocaloric devices.

Magnetoelastic coupling tends to drive magnetic phase transitions toward first-order behavior\cite{Bean_Rodbell1962}, thereby it can potentially enhance MCE. Magnetic materials with large magnetoelastic coupling may therefore show a strong magnetocaloric response. Bocarsly \emph{et al}. introduced a magnetocaloric descriptor based on the structural deformation between FM and nonmagnetic states\cite{bocarselydata} and applied it to characterize MCE in various magnetic materials. However, the nonmagnetic state is generally an inadequate representation of the PM state, which is characterized by disordered local magnetic moments.

In this work, we employ Landau theory to develop a magnetoelastic coupling descriptor capable of identifying magnetic materials near a first-order phase transition, where a strong MCE is expected. The descriptor can be calculated from \emph{ab initio} electronic structure calculations performed in the PM state, which we model using supercells with special quasi-random (collinear) spin configurations. 
We compute the descriptor for a range of known magnetocaloric materials and identify those in which the MCE is driven by magnetoelastic coupling. The descriptor is then applied in a HTP screening of the L1$_2$ family of compounds, predicting the cubic L1$_2$ phases of Mn$_3$Ge and Mn$_3$Sb as promising candidates with large MCE.

\section{Landau Theory of MCE}

We consider a magnetic material with magnetization (total magnetic moment per magnetic atom) $M$ subjected to a symmetry-conserving distortion. The distortion is described by a generalized strain $\bm{\epsilon}$, which is a symbolic vector whose elements represent, in general, changes in both the lattice parameters and the internal atomic coordinates. The Landau free energy functional (per magnetic atom) is given by
\begin{eqnarray}
G(M,\bm{\epsilon},T,H)=\frac{1}{2}A_0(T-T_C)M^2+\frac{1}{4}BM^4 -MH \nonumber \\
-\frac{1}{2}\mathbf{a}\cdot\bm{\epsilon} M^2+\frac{1}{2}\bm{\epsilon}\cdot\hat{K}^{-1}\cdot\bm{\epsilon}.
\label{Landau}
\end{eqnarray}
Here, $A_0$ and $B$ are positive temperature independent constants, $H$ is the magnetic field, $T$ is the temperature, and $T_C$ is the Curie temperature. The first three terms represent the well-known Landau free energy model used to describe magnetic phase transitions. The fourth term accounts for the magnetoelastic coupling, while the last term corresponds to the elastic energy. The symbolic vector $\mathbf{a}$ characterizes the strength of the magnetoelastic coupling, and $\hat{K}$ is the positive definite generalized compressibility tensor (per magnetic atom) associated with the considered distortion. The (equilibrium) free energy is given by the Landau free energy functional evaluated at the equilibrium values of $M$ and $\bm{\epsilon}$, which minimize the $G(M,\bm{\epsilon},T,H)$ function. The entropy can be then found from the partial derivative of the free energy with respect to temperature.

In the absence of magnetoelastic coupling ($\mathbf{a}=0$) and external stress, the equilibrium strain tensor vanishes. In this case, the isothermal entropy change (per magnetic atom) at $T=T_C$ due to the magnetic field changing from zero to a finite value $H$ can be expressed as follows:
\begin{equation}
\Delta S^0(T_C,H)=\frac{3^{1/3}}{2}k_B\left(\frac{M_sH}{k_BT_C}\right)^{2/3}
\label{DS0}
\end{equation}
where $M_s$ is the saturation magnetization. 

For a nonzero magnetoelastic coupling, the equilibrium strain tensor is given by
\begin{equation}
\bm{\epsilon}=\frac{1}{2}\hat{K}\cdot\mathbf{a}M^2
\end{equation}
Substituting this result into Eq.~(\ref{Landau}) yields the following expression for the Landau free energy
\begin{eqnarray}
 G(M,T,H)=\frac{1}{2}A_0(T-T_C)M^2+\frac{1}{4}BM^4-MH \nonumber \\
 -\frac{1}{8}\mathbf{a}\cdot\hat{K}\cdot\mathbf{a}M^4.
\label{Landau2}
\end{eqnarray}
The magnetoelastic coupling thus introduces an additional fourth order term in magnetization within the magnetic Landau free energy, which is negative. When the magnetoelastic coupling is strong, this negative contribution can outweigh the original positive fourth-order term (the second term in Eq.~(\ref{Landau2})), thereby rendering the magnetic phase transition first-order. This behavior aligns with the predictions of the Bean–Rodbell model\cite{Bean_Rodbell1962}, and suggests that magnetoelastic coupling enhances the temperature sensitivity of magnetization, thereby increasing the MCE.  Indeed, the following formula for the isothermal entropy change at $T_C$ can be derived
\begin{equation}
\Delta S(T_C,H)=\Delta S^0(T_C,H)\left(\frac{1}{1-\frac{4}{3}\eta}\right)^{2/3}.
\label{final}
\end{equation}
Here, $\eta$ is a positive parameter that represent the relative importance of the magnetoelastic coupling. It is given by
\begin{equation}
\eta=\frac{\frac{1}{8}\mathbf{a}\cdot\hat{K}\cdot\mathbf{a}M_s^4}{k_BT_C}=\frac{E^{\text{rel}}_{PM-FM}}{k_BT_C},
\label{eta}
\end{equation}
where $E^{\text{rel}}_{PM-FM}=\frac{1}{8}\mathbf{a}\cdot\hat{K}\cdot\mathbf{a}M_s^4$ represents the magnetic relaxation energy (per magnetic atom): the energy lowering due to structural relaxation of the PM atomic structure when the fully ordered FM state sets in. Within the Landau model, this quantity can be equivalently viwed as the difference in energy between the PM state, evaluated with the atomic structure of the fully ordered FM state, and the energy of the PM state with its own equilibrium atomic structure.

As evident from Eq.~(\ref{final}), a nonzero $\eta$ value increases the isothermal entropy change as compared to case without the magnetoelastic coupling. This enhancement becomes particularly significant as $\eta$ becomes close to the critical value of 0.75. At the limiting case of $\eta=0.75$ , the second and the fourth terms in Eq.~(\ref{Landau2}) exactly cancel, eliminating the fourth-order term in magnetization from the Landau free energy. This marks a tricritical point, where the nature of the magnetic phase transition shifts from second order to first order. For $\eta\geq0.75$, higher-order terms in magnetization must be incorporated into the Landau free energy functional, and therefore Eq.~(\ref{final}) is invalid. Nonetheless, in such cases, a strong MCE is still expected due to the first-order nature of the magnetic transition.

The parameter $\eta$ is related to the strain dependence of the Curie temperature. When a magnetic material is subjected to a fixed strain $\bm{\epsilon}$, the Landau free energy can be written as
\begin{eqnarray}
G(M,T,H)=\frac{1}{2}A_0(T-T_C-\frac{1}{A_0}\mathbf{a}\cdot\bm{\epsilon})M^2 \nonumber \\
+\frac{1}{4}BM^4-MH.  
\end{eqnarray}
Note that, in contrast to Eq.~(\ref{Landau}), $\bm{\epsilon}$ is treated as a fixed parameter rather than a variational variable in the Landau free energy. The temperature of the magnetic phase transition corresponds to the point at which the coefficient of the second-order term in magnetization vanishes. Thus, the strain-dependent Curie temperature, $T_C^\prime$, is given by
\begin{equation}
T_C^\prime=T_C+\frac{1}{A_0}\mathbf{a}\cdot\bm{\epsilon}=T_C\left(1+\bm{\beta}\cdot\bm{\epsilon}\right),
\end{equation}
where $T_C$ is the Curie temperature in the absence of external stress and the $\bm{\beta}$ symbolic vector describes the dependence of the Curie temperature on the strain. Using Eq.~(\ref{eta}) the following relation between $\eta$ and $\bm{\beta}$ can be derived
\begin{equation}
\eta=\frac{9}{8}k_BT_C\bm{\beta}\cdot\hat{K}\cdot\bm{\beta}.
\label{beta-eta}
\end{equation}
Based on the above discussion, the parameter $\eta$ quantifies the enhancement of the isothermal entropy change due to magnetoelastic coupling. In particular, it may serve as a useful descriptor in HTP searches for identifying magnetic materials where a strong magnetoelastic coupling brings the system on the brink of a first-order magnetic phase transition and thereby yielding a strong MCE.

\section{\emph{Ab Initio} Methodology} 

In order to employ the $\eta$ parameter as a magnetoelastic descriptor in HTP \emph{ab initio} screening for potential magnetocaloric materials, it is crucial that $\eta$ can be efficiently evaluated using electronic structure methods. According to the Eq.~(\ref{eta}), the key quantity needed for the calculations of $\eta$ is the magnetic relaxation energy that can be expressed as
\begin{equation}
E^\text{rel}_{PM-FM}=E_{PM}\big(\{\textbf{R}\}_{FM}\big)-E_{PM}\big(\{\textbf{R}\}_{PM}\big),
\label{Erel}
\end{equation}
where $E_{PM}\big(\{\textbf{R}\}_{FM}\big)$ and $E_{PM}\big(\{\textbf{R}\}_{PM}\big)$ denote the energies (per magnetic atom) of the system in the PM state, evaluated using atomic coordinates optimized at the FM and PM states, respectively. Note that $E_{PM}\big(\{\textbf{R}\}_{PM}\big)$ is always lower than $E_{PM}\big(\{\textbf{R}\}_{FM}\big)$, and thus the magnetic relaxation energy is a positive quantity.

\emph{Ab initio} calculations of the magnetoelastic descriptor require therefore structural relaxations in the PM state. However, accurately modeling the PM state from first principles presents a significant challenge. A straightforward approach to represent the PM state as nonmagnetic is justified only for strongly itinerant materials, in which magnetic moments vanish above the magnetic transition temperature. For majority of magnetic materials, however the PM state consists of disordered fluctuating magnetic moments, and its electronic structure differs significantly from that of the nonmagnetic state.

Inspired by single-site electronic structure theories of random alloys based on the coherent potential approximation (CPA) \cite{Soven1967,Velicky1969}, the disordered local moment (DLM) approach\cite{Gyorffy1985} has been developed to model the paramagnetic (PM) state. The DLM method has proven highly successful in describing the electronic structure of magnetic materials that are not too itinerant \cite{Staunton2014}. However, a key limitation of this technique is that it does not allow for structural relaxations.

The supercell approach provides a potential alternative. In this method, a large supercell is constructed containing many magnetic atoms, each with its magnetic moment assigned a fixed but random orientation. Such systems can be calculated using plane-wave–pseudopotential-based techniques, which allow for efficient atomic relaxation calculations. However, a potential drawback of this approach is its high computational cost: to adequately represent the paramagnetic state with random moment orientations, relatively large supercells are typically required. Moreover, noncollinear magnetic calculations are necessary, which are significantly more expensive computationally. These factors present a serious obstacle for HTP studies.

\begin{figure} [h]
\centering
\includegraphics[width=1.0\linewidth]{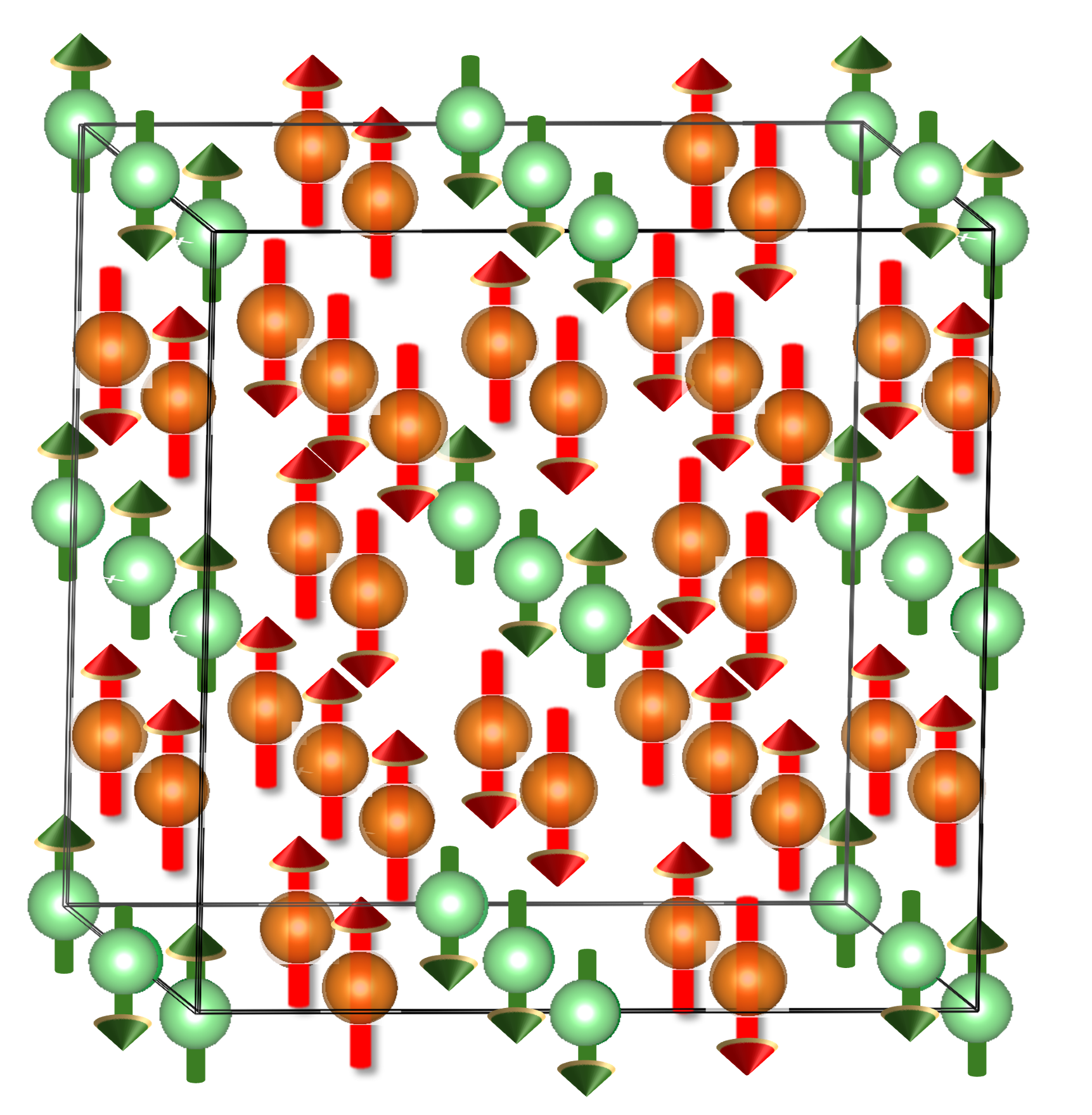}\vspace{-0pt}
\caption {SQS Ising PM spin configuration for a 2$\times$2$\times$2 supercell of the A$_3$B L1$_2$ structure. Both A (orange) and B (green) atoms are assumed to be magnetic.}
\label{fig: supercell}
\end{figure}

To improve the efficiency of the supercell approach, we draw on ideas from the theory of disordered alloys. Instead of assigning completely random moment orientations, we use special quasi-random structures (SQS) \cite{SQS1,SQS2} to assign the moment directions. This approach ensures that the correlation functions between the first few shells of neighbors vanish, allowing a significantly smaller supercell to accurately represent the disordered state. In this work, the SQS method is applied to Ising-like variables, where each magnetic moment is assigned either an up or down orientation (see Fig.~\ref{fig: supercell}). Such an Ising model of the paramagnetic state is motivated by the single-site CPA-DLM theory, in which the PM state is represented as a binary alloy of spin-up and spin-down local moments\cite{Gyorffy1985,Staunton2014}. Furthermore, this formulation allows the use of collinear magnetic calculations, making the approach considerably more computationally feasible.

For this approach to work, the magnetic moments must maintain their assigned orientations throughout the calculations. In more itinerant magnetic systems, however, the moments can often change direction during self-consistent calculations, making our approach unsuitable for relative energy evaluations. This issue tends to occur more frequently during structural relaxation calculations, whereas for fixed-ion (single-point) calculations, the moment orientations are typically preserved. Therefore, extending our method to estimate relative energies based on the forces and stresses obtained from single-point calculations could provide a practical and efficient improvement.

In addition to the magnetic relaxation energy, the Curie temperature is required for evaluating $\eta$. In many cases, the experimental value of $T_C$ can be used. However, for ferromagnetic materials lacking experimental data, $T_C$ can be estimated by assuming that their finite-temperature magnetic behavior is described by the classical Heisenberg Hamiltonian,
\begin{equation}
H=-\frac{1}{2}\sum_{ij}J_{ij}\hat{\mathbf{e}}_i\cdot\hat{\mathbf{e}}_j,
\end{equation}
where the summation runs over magnetic atoms, $J_{ij}$ denotes the exchange coupling between atoms $i$ and $j$, and $\hat{\mathbf{e}}_i$ is a unit vector along the direction of the magnetic moment at site $i$. Within the mean-field approximation (MFA), the Curie temperature is given by
\begin{equation}
k_B T_C = \lambda_{\max}(\mathbf{J}_0)/3,
\label{Tc_0}
\end{equation}
where $\lambda_{\max}(\mathbf{J}_0)$ is the largest eigenvalue of the $n \times n$ exchange matrix $\mathbf{J}_0$, with $n$ denoting the number of magnetic sublattices (or, equivalently, the number of distinct magnetic atom types). Its elements are defined as $(\mathbf{J}_0)_{\alpha\beta} = \sum_{j \in \beta} J_{i_\alpha j}$, where $i_\alpha$ is any magnetic atom of type $\alpha$, and the sum runs over magnetic atoms of type $\beta$. In the single-sublattice case, the largest eigenvalue is simply $J_0 = \sum_j J_{ij}$. In the MFA, $\lambda_{\max}(\mathbf{J}_0)$ corresponds to twice the energy difference between the random PM and perfectly ordered FM states. Consequently, the Curie temperature can be expressed as
\begin{equation}
k_BT_C=\frac{2}{3}\Big[E_{PM}\big(\{\textbf{R}\}_{FM}\big)-E_{FM}\big(\{\textbf{R}\}_{FM}\big)\Big],
\label{Tc}
\end{equation}
where \(E_{FM}(\{\mathbf{R}\}_{FM})\) is the energy (per magnetic atom) of the system in the fully ordered ferromagnetic state, evaluated using atomic coordinates optimized for this configuration.

The initial crystal structures of the studied materials were obtained from the Materials Project database \cite{MatProj}. These structures were subsequently symmetrized using the \texttt{spglib} library\cite{spglib,spglibv2} and the \texttt{pymatgen} Python package. Structural optimizations and total energy calculations were performed for both the fully ordered FM and PM states within the framework of density functional theory (DFT) \cite{Kohn-Sham}. The Kohn–Sham equations were solved using the projector augmented-wave method \cite{BlochlPRBPAW, KressePRBUSPP}, as implemented in the Vienna \emph{ab initio} simulation package (VASP) \cite{VASP4}. The PM state was modeled using supercells with disordered collinear spin configurations generated via the Monte Carlo special quasirandom structure (MC-SQS) technique \cite{VanDeWalle2013}, ensuring that the spin–spin correlation functions vanish for the first few neighbor shells. The MC-SQS simulations were carried out using the \texttt{mcsqs} module of the ATAT software package \cite{ATAT}.

Calculations were performed for a set of known magnetocaloric materials and for magnetic $L1_2$ compounds. For the first group we used the PBEsol exchange-correlation functional \cite{PBEsol}. The only exception was Gd, for which we used LDA+U with the parameters reported in Ref.~\citenum{Kurz2002}. Full structural relaxations, including optimization of the cell volume, shape, and atomic positions, were performed until the Hellmann–Feynman forces on all atoms were less than 0.01 eV/Å. The $k$-point meshes were chosen to ensure convergence of the total energy of the FM state within 1 meV. The resulting $k$-point meshes for different compounds are listed in Table~S1 in the Supplemental Materials. For the PM calculations, the supercell sizes used for each compound are also given in Table S1.

For the L1$_2$ compounds, the PBE exchange-correlation functional was used \cite{PBE}. For the FM state, the primitive unit cell of the L1$_2$ structure was employed, whereas for the PM state the corresponding $2\times2\times2$ supercell was used. A $12\times12\times12$ $k$-point mesh was used for the FM calculations, and the mesh was scaled accordingly for the larger PM supercell calculations. For most of the compounds, the same SQS-PM spin configuration was adopted. It was generated using MC-SQS simulations by imposing the condition that all spin-spin correlation functions vanish for the first four shells of nearest neighbors. The resulting SQS spin configuration is shown in Fig.~\ref{fig: supercell} (for nonmagnetic atomic species, the spin orientation is irrelevant). While the particular SQS configuration in the supercell breaks the L1$_2$ cubic symmetry, the statistical ensemble average retains cubic symmetry, making the volume the only structural degree of freedom. Therefore, the structural optimizations were performed by freezing the internal degrees of freedom and relaxing only the volume in both the FM and PM states. For some L1$_2$ compounds, the imposed SQS-PM spin configuration was not preserved during structural relaxation. In such cases, the equilibrium PM volume was determined by performing a series of fixed-volume calculations and fitting the resulting total energies to a parabolic equation of state. In a few cases (Fe$_3$Ni, Fe$_3$Pd, Co$_3$Fe, and Pt$_3$Co), the spin configuration was not preserved even in the fixed-volume calculations, and for these materials we used a different SQS-PM spin configuration in which only the first two shells of neighbors were required to vanish. For the other compounds, the results do not depend significantly on which of the two SQS spin configurations is used.

\section{Results}

\subsection{Calculations of the Magnetoelastic Descriptor for Known Magnetocalorics}

To assess the relevance of the proposed magnetoelastic descriptor in real systems, we evaluated it for a set of experimentally known magnetocaloric materials. The materials considered include the ordered compounds compiled by Bocarsly \emph{et al.}\cite{bocarselydata}, which encompass a diverse range of rare-earth free magnetocalorics, as well as elemental gadolinium (Gd), a benchmark material for magnetocaloric performance near room temperature. By applying the descriptor to this dataset, we aim to reveal how the strength of magnetoelastic coupling correlates with the observed MCE, particularly the magnitude of the entropy change and the character of the magnetic transition. This analysis allows one to distinguish systems in which a strong MCE is primarily driven by magnetoelastic coupling. 

\begin{figure}[h]
\centering
\includegraphics[width=1.0\linewidth]{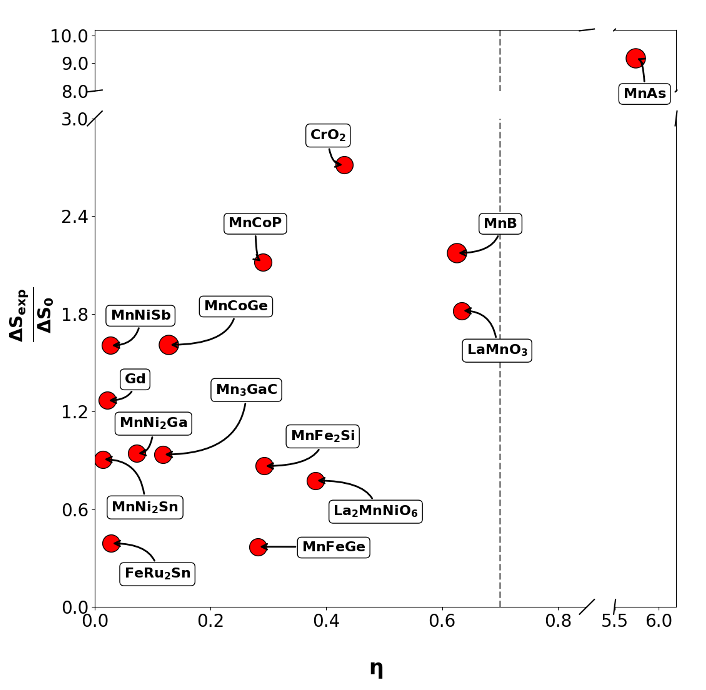}\vspace{-8pt}\
\caption{Experimental peak isothermal entropy change for an applied magnetic field of $\mu_0H = 2~\mathrm{T}$ as a function of the magnetoelastic descriptor $\eta$ for a set of known magnetocaloric materials. The experimental data are taken from Ref.~\cite{bocarselydata} and normalized by $\Delta S^0$ (Eq.~\ref{DS0}), evaluated using the experimental values of the saturation magnetization and Curie temperature. The descriptor $\eta$ was obtained from the calculated magnetic relaxation energy and the experimental $T_C$. The dashed vertical line denotes the critical threshold, $\eta = 0.75$, above which a first-order magnetic transition can be expected.}
\label{fig:SQS_Bocarsely_FM}
\end{figure}

Figure~\ref{fig:SQS_Bocarsely_FM} presents the experimental peak isothermal entropy change as a function of the magnetoelastic descriptor, $\eta$, for the considered  set of magnetocaloric materials. The descriptor $\eta$ was obtained using the calculated magnetic relaxation energy together with the experimental Curie temperature. It should be noted that the compilation by Bocarsly \emph{et al.} \cite{bocarselydata} contains a number of relatively itinerant systems, for which the local-moment description underlying our approach is less robust. As a result, 14 of the ordered compounds are not included in Fig.~\ref{fig:SQS_Bocarsely_FM}, because their SQS spin configurations were not preserved during structural relaxation. Consequently, $\eta$ could not be evaluated within our approach for these compounds.

For MnAs, the calculations were performed in the low-temperature NiAs-type phase and yield a very large descriptor value ($\eta=5.7$), far above the critical threshold of 0.75. This large value reflects a strong volume contraction (about 8\%) in the SQS-PM state, which is partly associated with a substantial reduction of the average magnetic moment from 2.52 in the FM state to 1.66 in the SQS-PM state. While the inclusion of longitudinal spin fluctuations would likely reduce the magnitude of the contraction, the magnetoelastic effect is expected to remain very strong. Experimentally, MnAs undergoes a first-order magnetostructural transition in which the low-temperature FM hexagonal NiAs-type phase transforms into a high-temperature PM orthorhombic MnP-type phase \cite{MnAs}, and a giant isothermal entropy change can be obtained \cite{MnAs2}. Previous work has shown that this transition is driven by a soft $M$-point phonon mode that becomes unstable upon volume contraction\cite{Lazewski2010}. In our calculations, we do not search for the competing structural phase, and thus the system remains in the hexagonal phase even in the SQS-PM state. However, we explicitly demonstrate that spin disorder is responsible for the volume contraction required to destabilize the $M$-point phonon mode. Interestingly, our calculations indicate that even in the absence of the structural instability, the magnetic transition would still be expected to be first order due to the strong magnetoelastic coupling.

Large descriptor values can arise even in the absence of structural instabilities. In particular, MnB has $\eta=0.62$. This indicates strong magnetoelastic coupling, which can enhance the magnetocaloric response and drive the magnetic transition toward first-order behavior. This interpretation is consistent with the reported weakly first-order magnetic transition and large isothermal entropy change \cite{Bocarsly2019}. A similar descriptor value ($\eta=0.63$) was obtained for LaMnO$_3$ perovskite (space group Pnma) assuming a FM ground state. Whereas stoichiometric LaMnO$_3$ is an antiferromagnet, ferromagnetism can be stabilized experimentally by hole doping, most commonly through A-site substitution with alkaline-earth ions \cite{Zhou2015} or by oxygen nonstoichiometry \cite{Ritter1997}. The large value of $\eta$ suggests strong magnetoelastic coupling, reflecting the sensitivity of the superexchange interaction to local structural distortions that modify the Mn--O--Mn bond geometry. This is consistent with the sizable magnetocaloric response reported in doped LaMnO$_3$-based compounds \cite{Phan2007}, including cases where the FM--PM transition becomes first order\cite{Adams2004,Phan2007}.

Several materials exhibit intermediate descriptor values ($0.28 \lesssim \eta \lesssim 0.43$). This group includes rutile CrO$_2$, the double perovskite La$_2$MnNiO$_6$, Heusler alloy MnFe$_2$Si, and the MM$'$X compounds MnCoP and MnFeGe. For these systems, the magnetic exchange is also sensitive to changes in local geometry, giving rise to appreciable magnetoelastic coupling. Although this coupling is insufficient to drive the magnetic transition first order, it can still substantially enhance the magnetocaloric response. A large $\Delta S(T_C,H)/\Delta S^0(T_C,H)$ ratio is indeed observed for CrO$_2$ and MnCoP. For La$_2$MnNiO$_6$, MnFe$_2$Si, and MnFeGe, however, the ratio is smaller than one despite the significant $\eta$. This is not surprising, since $\Delta S^0(T_C,H)$ is only a simple mean-field expression and is not expected to capture all material-specific aspects of the bare magnetocaloric response, particularly in low-coordination lattices.

Finally, we have materials with low descriptor values ($\eta < 0.12$). This set includes the half-Heusler MnNiSb, Gd, the anti-perovskite Mn$_3$GaC, MM$'$X compound MnCoGe, and the Heusler compounds MnNi$_2$Sn, MnNi$_2$Ga, and FeRu$_2$Sn. For these systems, the magnetoelastic contribution to the magnetocaloric response is expected to be small.

\subsection{HTP search for $L1_2$ magnetocalorics}

\begin{figure} [h]
\centering
\includegraphics[width=1.0\linewidth]{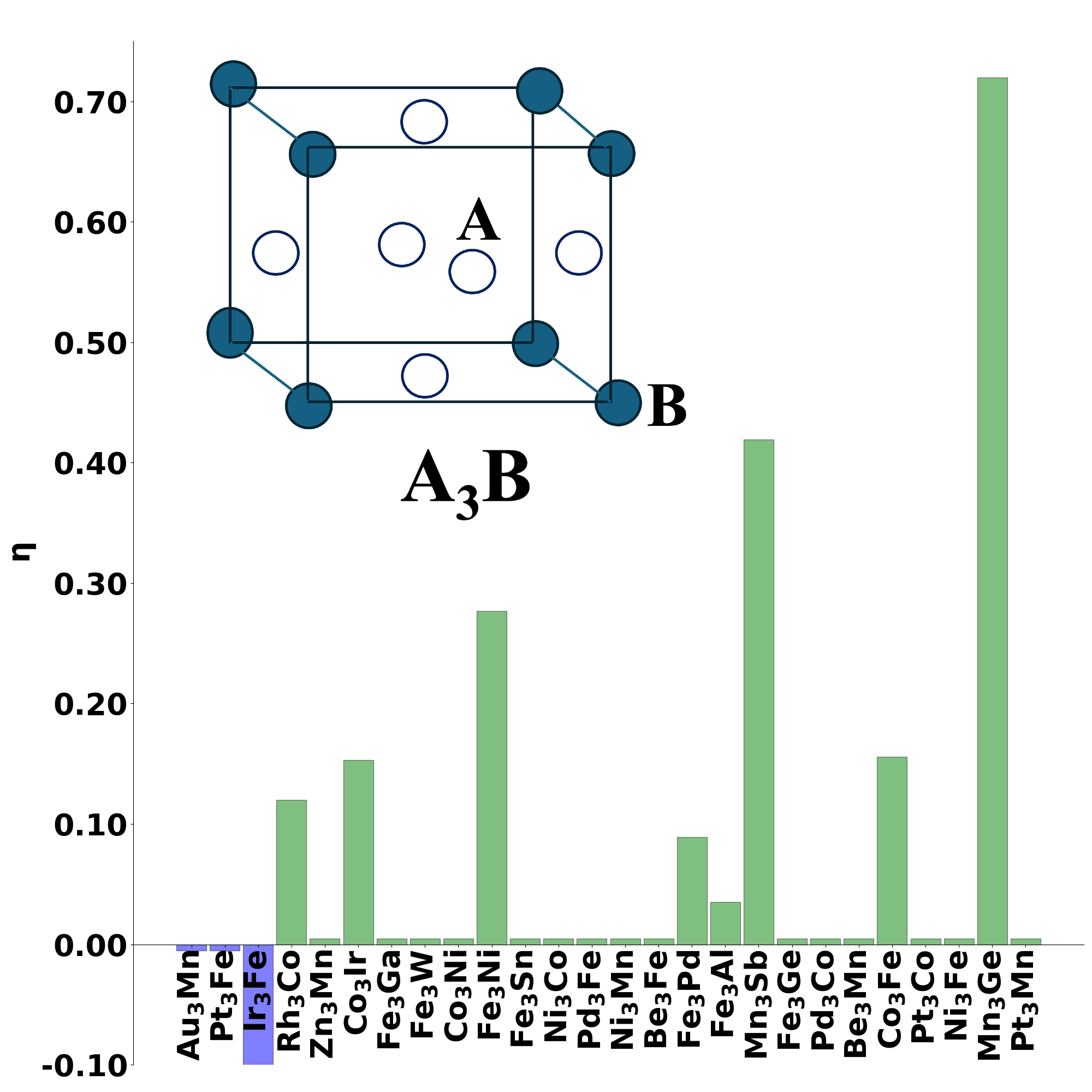}\vspace{-5pt}
\caption {Magnetoelastic descriptor $\eta$ evaluated for a set of magnetic L1$_2$ compounds. The descriptor was obtained from the calculated magnetic relaxation energy and $T_C$ computed from Eq.~\ref{Tc}. Negative values of $\eta$ arise from negative $T_C$, indicating that the FM state is not the magnetic ground state. The inset shows the crystal structure of $A_3B$ cubic L1$_2$ compounds.}
\label{fig: Eta_L12}
\end{figure}

To further explore the predictive power of the descriptor $\eta$, we apply it in a HTP screening study of the $L1_2$ family of intermetallic compounds in search of promising magnetocaloric materials exhibiting strong magnetoelastic coupling. The $L1_2$ crystal structure, illustrated in the inset of Fig.~\ref{fig: Eta_L12}, is a cubic ordered structure belonging to the $Pm\bar{3}m$ space group. The $A_3B$ compounds crystallizing in the $L1_2$ structure have the $A$ atoms occupying the face-centered positions, while the $B$ atoms are located at the cube corners. The high structural symmetry of the $L1_2$ phase results in only a single structural degree of freedom (the volume), which simplifies the analysis of magnetoelastic coupling. At the same time, $L1_2$ compounds exhibit substantial chemical flexibility, giving rise to a significant compound space well suited for HTP screening.

The initial set of candidate $L1_2$ compounds was generated by querying the Materials Project database \cite{MatProj}. Since magnetocaloric applications require FM materials with Curie temperatures near room temperature, the database search was restricted to compounds containing Mn, Fe, or Co and exhibiting a total magnetic moment exceeding $0.5~\mu_{\mathrm{B}}$ per formula unit. To focus on materials with greater practical relevance and improved experimental accessibility, compounds containing rare-earth, actinide, or radioactive elements were excluded from consideration. The query included both experimentally reported materials and theoretically predicted compounds available in the database. For theoretically predicted compounds, only systems with an energy above the convex hull not exceeding 0.1~eV/atom were retained in order to ensure that the compounds remain thermodynamically accessible. After applying these selection criteria, the initial pool of candidates was reduced to 31 $L1_2$ compounds selected for further \emph{ab initio} screening.

For five of the selected $L1_2$ compounds, the FM state is not stable, and these materials were therefore discarded. For the remaining 26 compounds, the magnetoelastic descriptor was evaluated using the calculated magnetic relaxation energy and Curie temperature obtained from Eq.~\ref{Tc}. The results are shown in Fig.~\ref{fig: Eta_L12} and Table~S2 in the Supplemental Materials, where the numerical values of $\eta$ and the calculated $T_C$ are reported. There were no significant changes in the results when different SQS configurations were used. For a few compounds, the calculated $T_C$ is negative, which results in a negative descriptor value. This indicates that the FM state is not the magnetic ground state for these compounds, and thus they are not magnetocaloric candidates. 

For most of the considered materials, $\eta$ is very low, so a significant magnetoelastic contribution to the MCE is not expected. Notable exceptions are the Mn$_3$Ge and Mn$_3$Sb compounds. For Mn$_3$Ge, $\eta = 0.72$, which is close to the critical threshold of 0.75, indicating proximity to a first-order FM transition and a strongly enhanced magnetocaloric response. The large descriptor value is reflected in a significant volume enhancement (about 4\%) in the SQS-PM state relative to the FM state. Interestingly, spin disorder also increases the magnitude of the Mn local magnetic moment, from $1.78\mu_B$ in the FM state to $2.33\mu_B$ in the SQS-PM state. While the L1$_2$ phase of Mn$_3$Ge is metastable, it has been synthesized experimentally under high pressure, where FM behavior with $T_C \approx 400$ K has been observed \cite{Mn3Ge}. The calculated Curie temperature, $T_C \approx 150$ K, is much lower than the experimental value. This discrepancy may be caused by non-Heisenberg magnetic interactions that are not accounted for by Eq.~\ref{Tc}. Another possible explanation is the reduced degree of L1$_2$ ordering in the experimental sample.

Mn$_3$Sb has a smaller descriptor value ($\eta = 0.42$), even though the volume expansion upon transitioning to the SQS-PM state relative to the FM state is larger, at about 5\%. As in Mn$_3$Ge, spin disorder increases the magnitude of the Mn local magnetic moment, from $2.03\mu_B$ in the FM state to $2.73\mu_B$ in the SQS-PM state. The calculated Curie temperature is $T_C \approx 477$ K, substantially higher than that of Mn$_3$Ge. The L1$_2$ phase of Mn$_3$Sb has been prepared experimentally, but its magnetic character remains debated, with one report suggesting weak ferromagnetism\cite{Yamashita2003} and another ferrimagnetism\cite{Ryzhkovskii2011}. The Mn$_{3}$Ge$_x$Sb$_{1-x}$ alloy may therefore offer an interesting route to tune magnetic interactions, magnetoelastic coupling, and the magnetocaloric response.

\section{Conclusions}

In this work, we derived a computable magnetoelastic descriptor for magnetocaloric materials from Landau theory of phase transitions. The descriptor measures the coupling between PM spin disorder and lattice relaxation, and it provides a dimensionless control parameter for the crossover from second-order to first-order magnetic behavior. It therefore predicts when magnetoelastic effects may drive a magnetic transition toward first-order character and produce a strong magnetocaloric response.

Although it targets the same underlying physics, the descriptor differs substantially from the computational proxy introduced in Ref.~\citenum{bocarselydata}, which measures the structural difference between the FM and nonmagnetic states \cite{bocarselydata}. By contrast, our descriptor compares the magnetic relaxation energy associated with the PM state to the magnetic energy scale itself, and is therefore more directly connected to the actual FM--PM phase transition. Grounded in Landau theory, the descriptor provides a natural criterion for whether the transition is expected to remain second order or become first order, and it quantifies the enhancement of the magnetocaloric response due to magnetoelastic coupling. In the present implementation, we model the PM state using disordered local magnetic moments represented by special quasi-random collinear spin supercells. This approach is computationally more demanding but, for the majority of magnetic materials, provides a more realistic description of the PM state than the nonmagnetic approximation. Importantly, the descriptor itself is not tied to this particular representation of the PM state and can, in principle, be evaluated using other suitable approaches for modeling the PM state.

We applied the descriptor to a set of known magnetocaloric materials to provide a clear measure of the magnetoelastic contribution to the magnetocaloric effect and to identify systems in which magnetoelastic coupling brings the material to the brink of a first-order phase transition. A giant magnetovolume effect was found for MnAs, and proximity to a first-order magnetic transition was identified for MnB and LaMnO$_3$, in agreement with experimental reports.

The descriptor was also used in a high-throughput \emph{ab initio} screening of magnetic L1$_2$ compounds. The screening identified candidate materials with strong magnetoelastic coupling and enhanced magnetocaloric response. In particular, the L1$_2$ cubic phases of Mn$_3$Ge and Mn$_3$Sb emerge as especially promising candidates: Mn$_3$Ge shows a descriptor value close to the critical threshold associated with first-order magnetic behavior, while Mn$_3$Sb exhibits a smaller but still sizable descriptor value. These results demonstrate that the descriptor can serve as an efficient computational filter for discovering new magnetocaloric materials.

\begin{acknowledgements}
This work was supported by the U.S. Department of Energy (DOE) Established Program to Stimulate Competitive Research (EPSCoR) grant no. DE-SC0024284.
\end{acknowledgements}

\bibliographystyle{unsrt}

\bibliography{bibilo.bib}

\end{document}